\documentclass[aps,prl,twocolumn,showpacs]{revtex4-1}
\usepackage{amsmath,amssymb,bm}
\usepackage{dsfont}
\usepackage{graphicx}

\usepackage{color}
\newcommand{\bra}[1]{\langle {#1} |}
\newcommand{\ket}[1]{| {#1} \rangle}
\newcommand{\phih}{\hat{\phi}}
\newcommand{\Phihb}{\hat{\bm \Phi}}
\newcommand{\phihb}{\hat{\bm \phi}}

\newcommand{\Phih}{\hat{ \Phi}}

\newcommand{\psib}{{\bm \psi}}
\newcommand{\Hh}{\hat{{\cal H}}}
\newcommand{\latvec}{{\bf a}}
\newcommand{\ah}{\hat{\alpha}}
\newcommand{\shift}{\Delta}

\makeatletter
\@ifundefined{definecolor}
 {\usepackage{color}}{}
\usepackage[colorlinks=true,linkcolor=blue,urlcolor=blue,citecolor=blue]{hyperref}

\begin{document}
\title{Selective Population of Edge States in a 2D Topological Band System}
\date{\today}
\author{Bogdan Galilo,$^1$ Derek K.\ K.\  Lee,$^2$ and Ryan Barnett$^1$}
\affiliation{$^1$Department of Mathematics and $^2$Blackett Laboratory, Imperial College London,
London SW7 2AZ, United Kingdom}
\begin{abstract}
We consider a system of interacting spin-one atoms in a hexagonal lattice under the presence of a synthetic gauge field.
Quenching the quadratic Zeeman field is shown to lead to a dynamical instability of the edge modes.  This, in turn, leads
to a spin current along the boundary of the system which grows exponentially fast in time following the quench.  
Tuning the magnitude of the quench can be used to selectively populate edge modes of different momenta.
Implications of the intrinsic symmetries of Hamiltonian on the dynamics are discussed.  The results hold for atoms
with both antiferromagnetic and ferromagnetic interactions.
\end{abstract}
\maketitle

Topological Bloch bands and their concomitant protected edge modes
play important physical 
roles in several solid-state materials including quantum Hall systems
\cite{thouless82}
and  topological insulators/superconductors \cite{hasan10,qi11}.
Recent years have experienced remarkable progress in engineering
systems which possess topological Bloch bands as a result of induced
effective gauge fields.  Effective gauge fields have been
realized in systems of ultracold atoms through mechanical rotation,
optical lattice `shaking', and laser-assisted tunneling (see
\cite{goldman2014} for a review).  Recent milestones include the
experimental realization of the Rice-Mele \cite{atala13}, Hofstadter
\cite{williams10, aidelsburger13,miyake13}, and Haldane \cite{jotzu14}
models in optical lattice systems.  
The emerging field of topological photonic lattices offers a separate
and complementary system where topological Bloch bands have also
recently been realized \cite{haldane08, wang09,hafezi11,rechtsman13}.
 Most of the experimental systems with ultracold atoms
involve temperatures for which the particle dynamics can be
accurately described by non-interacting theories.  However, as
experimental techniques are refined and quantum degeneracy is reached,
interactions and particle statistics will play an essential role in
the physics of these systems.

Of central interest in each of these models is the presence of
topologically protected edge modes.  The physical consequence of edge modes
for a degenerate Fermi gas in a topological band is clear: if the
Fermi energy resides within the bulk band gap, the near-equilibrium
dynamics will be described by degrees of freedom localized along the
edge of the system.  On the other hand, for bosonic systems, 
particles will generally
condense into the lowest bulk band, leaving the higher-energy edge
states largely unimportant for the dynamics of the system. 

In this work, we propose a scheme to bring to the fore the role of
edge states in the dynamics of bosons in a prototypical
two-dimensional topological band system.  In particular, we describe
how a quantum quench of an interacting spinor generalization of the
well-known Kane-Mele model \cite{kane05} can result in an
exponentially fast growth in the population of edge (but not bulk)
states. This will be exhibited as an exponential growth of the spin
current along the boundary of the system. Furthermore, the momenta of
these edge states can be selected by tuning the magnitude of the
quench.

Before proceeding, we briefly comment on previous related work.
Probing the topology of band systems through a quench has received
considerable attention in the recent theoretical literature
(\emph{e.g.} \cite{killi12,foster13,
  kells14,hauke14,vasseur14,sacramento14, eriksson15,caio15}). One of
us \cite{barnett13} has described the exponentially fast population of
an edge mode in perhaps the simplest topological system, the
Su-Schrieffer-Heeger (SSH) model \cite{heeger88}, by preparing a
bosonic gas in an excited spatial mode. A photonic version of the SSH model was
in fact realized \cite{poli15} where on-site
absorption was shown to lead to fast population of edge states
\cite{schomerus13}.  Quadratic fermionic hamiltonians have been
previously classified in terms of their symmetries
\cite{schnyder08,kitaev09}.  Much less work, however, has been carried
out for bosonic systems.  An expression for the Chern number,
accounting for the symplectic transformation properties of bosonic
systems was obtained in \cite{shindou13}.  In \cite{engelhardt15}, the
topology of real Bogoliubov excitations in inversion-symmetric
lattices was analyzed.  However, to our knowledge, a classification of
quadratic bosonic systems allowing for dynamical instabilities does
not presently exist.

The Haldane model \cite{haldane88} was realized in an ultracold atomic
system by the Esslinger group \cite{jotzu14}.  This system involves atoms
on a hexagonal optical lattice where spin-orbit coupling is induced
through shaking.  In this work, we consider a spin-one version of this
system where atoms experience spin-orbit coupling proportional to the
$z$-component of their spins.  In particular, we introduce the
following generalization of the Kane-Mele model \cite{kane05}:
\begin{align}
\label{S1KM}
\Hh_{\rm S1KM}=-w \sum_{\langle ij \rangle} \Phihb_i^\dagger \Phihb_j + i \lambda
\sum_{\langle \langle ij \rangle \rangle} \nu_{ij} \Phihb_i^\dagger  S_z \Phihb_j
\end{align}
where $\Phihb_i=( \Phih_{i,1}, \Phih_{i,0},\Phih_{i,-1})^T$ is a
vector composed of bosonic annihilation operators at site $i$ for each
spin component and we denote the $3 \times 3$ spin-one matrices as
${\bf S} = (S_x, S_y, S_z)$.  The second term above describes hopping
between second neighbors, and $\nu_{ij}=+1 (-1)$ if the atom makes a
left (right) turn to reach a second-neighbor site \cite{kane05}.  The
spin components are decoupled in $\Hh_{\rm S1KM}$: the spin-zero
component is described by the nearest-neighbor graphene model with
hopping $w$ while the spin-$\pm 1$ components are described by two
Haldane models with opposite magnetic fields. We restrict our
attention to $w>\sqrt{3} |\lambda|$ so that the single-particle state
of lowest energy occurs at the centre of the Brillouin zone.

Next, we include on-site interactions that preserve spin rotation
invariance \cite{ho98, ohmi98, stamperkurn13}:
\begin{align}
\Hh_{\rm int} = \sum_i \left[ \frac{U}{2} \left(\Phihb_i^\dagger
    \Phihb_i \right)^2+ \frac{U_s}{2}  \left(\Phihb_i^\dagger {\bf S}
  \Phihb_i\right)^2 \right]
\end{align}
where $U$ and $U_s$ describe the magnitude of the density and spin
interactions respectively.  Finally, we introduce the standard
quadratic Zeeman effect for spinor condensates \cite{stamperkurn13}:
$ \Hh_{\rm ext} = q \sum_i \Phihb_i^\dagger \left(S_z\right)^2
\Phihb_i $.
While external magnetic fields will provide only positive values of
$q$, microwave fields can be utilized to access both positive and
negative quadratic Zeeman shifts \cite{gerbier06}.  Introducing a
chemical potential $\mu$, the full Hamiltonian reads
$ \Hh=\Hh_{\rm S1KM} + \Hh_{\rm int} + \Hh_{\rm ext} - \mu \sum_i
\Phihb_i^\dagger \Phihb_i.  $
which is invariant under time reversal and global spin
rotations about the $z$-axis, the importance of which will be
addressed below.

In this work, we consider a quantum quench that abruptly
changes the quadratic Zeeman energy $q$ from an initially large and
positive value to a final value $q_f$.  This form of quenching has been
experimentally achieved in several experiments in the past decade (see
\cite{stamperkurn13} and references therein).  The initial
state is a coherent state with all bosons in a spatially
uniform spin-zero state:
\begin{align}
\label{psiinitial}
\ket{\Psi_{\rm in}} = e^{-\frac{1}{2}N_p} e^{\sqrt{\bar{n}} \sum_i \Phih_{i,0}^\dagger} \ket{0}
\end{align}
where $N_p$ is the total atom number. There are no particles with
$s_z=\pm 1$ in this state. This initial state is the
mean-field superfluid ground state of the graphene-lattice boson
Hubbard model. A variational calculation shows that the
average number of bosons per site, $\bar{n}$, is related to the
chemical potential by  $\mu=\bar{n}U -3w$.  

To investigate the ensuing dynamics after the quench, we consider small
fluctuations of the Hamiltonian with $q=q_f$ around the initial state
\eqref{psiinitial}.
Let $\phihb_i= \Phihb_i - (0,\sqrt{\bar{n}},0)^T$ where
$\phihb_i=(\phih_{i,1},\phih_{i,0},\phih_{i,-1})^T$.
The Hamiltonian can be expanded to quadratic order in $\phihb_i$ as
$\Hh= \bra{\Psi_{\rm in}}\Hh\ket{\Psi_{\rm in}}+\Hh_B$.  One finds
\begin{align}
\label{Eq:bog}
\Hh_B
&= -w \sum_{\langle ij \rangle} \phihb_i^\dagger \phihb_j + i \lambda
\!\sum_{\langle \langle ij \rangle \rangle} \nu_{ij} \phihb_i^\dagger
S_z \phihb_j   + \sum_i \phihb_i^\dagger M \phihb_i
\notag
\\ &+
\sum_i \left[ \left(
\frac{U\bar{n}}{2} \phih_{i,0} \phih_{i,0} + U_s \bar{n}
\phih_{i,1} \phih_{i,-1}\right) + {\rm H.c.}
\right] 
\end{align}
where
$M={\rm diag}(3w+U_s\bar{n} +q_f, 3w+U\bar{n},3w+U_s\bar{n}+q_f)$ and
$q_f$ is the quadratic Zeeman energy after the quench.  The
$S_z$-rotation symmetry of $\Hh_B$ ensures that the spin-$\pm1$
components are decoupled from the spin-$0$ components and hence the
(\ref{Eq:bog}) can be written as $\Hh_B = \Hh_{0} + \Hh_{\pm1}$.  The
Hamiltonian $\Hh_0$ describes the dynamics of the spin-zero components
and has no spin-orbit coupling. It is readily diagonalized by
a Bogoliubov transformation.  The resulting spectrum is stable and
exhibits the usual linearly-dispersing phonon mode. From now on, we
will focus on the spin-$\pm 1$ sector described by $\Hh_{\pm 1}$.

As we are primarily interested in the dynamics of the edge states of
this model, we will focus on the strip geometry.  Denoting the
primitive lattice vectors of graphene as $\latvec_1$ and $\latvec_2$, 
we consider open (periodic) boundary
conditions along the $\latvec_1$ ($\latvec_2$) direction.  It is
instructive to rewrite (\ref{Eq:bog}) in the eigenbasis of the
non-interacting spin-one Kane-Mele model (\ref{S1KM}) in this
geometry. The spin-$\pm 1$ Hamiltonian becomes
\begin{align}
\Hh_{\pm1} = \sum_{k,\nu}  & \left[ ( \varepsilon^{(\nu)}_{k} - \shift) 
(\ah_{k,\nu,1}^\dagger \ah_{k,\nu,1}  + \ah_{-k,\nu,-1}^\dagger \ah_{-k,\nu,-1} )
\right. \notag\\
& \left. +\ U_s \bar{n} (\ah_{k,\nu,1} \ah_{-k,\nu,-1} + {\rm H.c.}) \right].
\end{align}
Here, $\varepsilon_k^{(\nu)}$ are the single-particle energies of the Haldane model in the strip geometry,
$k={\bf k}\cdot \latvec_2$ is the momentum along the periodic
direction, and $\ah_{k,\nu,m}$ annihilates a boson in the eigenbasis
of (\ref{S1KM}).  We have introduced $\shift=-U_s \bar{n} -3w - q_f$
which serves as the tuning parameter for our quench.  
Due to time-reversal symmetry and the spatial
uniformity  of the initial state, 
$\Hh_{\pm1}$ can be separated into pairwise couplings
between ($k,\nu,m$) and ($-k,\nu,-m$) modes 
which greatly simplifies the analysis.

Unlike $\Hh_0$, $\Hh_{\pm1}$ cannot in general be brought to diagonal
form and may exhibit a dynamical instability.  
Therefore, we focus instead on the Heisenberg equations of motion:
  $i \partial_t \ah_{k,\nu,\pm1}(t) = [\ah_{k,\nu,\pm1}(t), \Hh_{\pm1}]$
where $\ah_{k,\nu,\pm1}(t) = e^{i \Hh_{\pm 1} t } \ah_{k,\nu,\pm1} e^{-i \Hh_{\pm 1} t }$ and we have set $\hbar=1$.
These can be solved to give 
\begin{equation}
\label{EOM}
\begin{split}
\ah_{k,\nu,1} (t) &= A_{k,\nu}(t)  \ah_{k,\nu,1} + B_{k,\nu}(t) \ah_{-k,\nu,-1}^\dagger\,, \\
\ah_{-k,\nu,-1}(t) &= B_{k,\nu}(t) \ah_{k,\nu,1}^\dagger + A_{k,\nu}(t) \ah_{-k,\nu,-1}\,.
\end{split}
\end{equation}
where
$
A_{k,\nu} (t)=\cos(E_k^{(\nu)} t ) -i
(\varepsilon_k^{(\nu)}-\shift)
\sin(E_{k}^{(\nu)}t)/E_{k}^{(\nu)}
$
and
$
B_{k,\nu} (t) = -iU_s
  \bar{n}\sin(E_{k}^{(\nu)}t)/E_{k}^{(\nu)}
$
with the Bogoliubov energies
\begin{align}
\label{BogEn}
E_k^{(\nu)} = \sqrt{(\varepsilon_k^{(\nu)} - \Delta)^2- (U_s \bar{n})^2}.
\end{align}
For sufficiently low condensate depletion, the Bogoliubov Hamiltonian
can be used to propagate the initial 
state (\ref{psiinitial})  as \ $\ket{\Psi(t)}= e^{-i \Hh_B t}
\ket{\Psi_{\rm in}}$.   The solutions (\ref{EOM})  can then be used to obtain the expectation value of bilinear operators:
\begin{align}
\label{bilinear}
&\bra{\Psi(t)} \ah_{k,\nu,m}^\dagger \ah_{k',\nu',m'} \ket{\Psi(t)} = \delta_{k,k'} \delta_{\nu,\nu'} \delta_{m,m'} |B_{k,\nu}|^2 \\
&\bra{\Psi(t)} \ah_{k,\nu,m} \ah_{k',\nu',m'} \ket{\Psi(t)} = \delta_{k,-k'} \delta_{\nu,\nu'} \delta_{m,-m'} A_{k,\nu} B_{k,\nu}
\notag
\end{align}
for spin $m=\pm 1$ components.

\begin{figure}
 \includegraphics[width=85mm,angle=0]{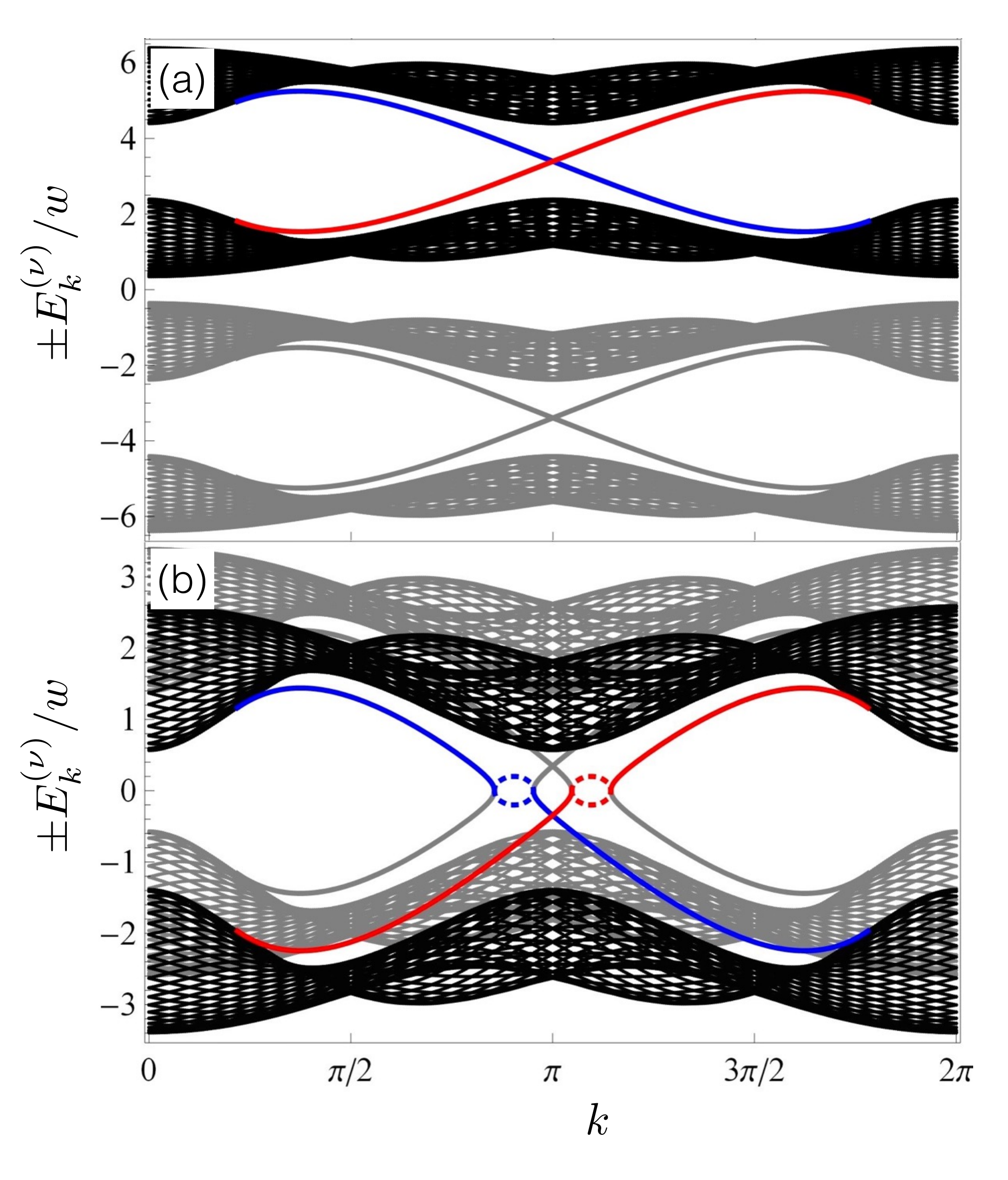}
 \caption{(Color Online)  
The Bogoliubov energy spectrum $\pm E_k^{(\nu)}$ corresponding to
(\ref{BogEn}) for the interacting spin-one Kane-Mele model in the
strip geometry.  Parameters in (a) are $U_s \bar n=0.2 w$,
$\lambda=w/2$ and $q_f=U_s \bar{n}$ which corresponds to a shallow
quench with stable spectrum.  Parameters in (b) are the same as (a)
except $q_f=-3w-3 U_s \bar{n}$ for which the bulk states are stable
while the edge states experience an exponentially fast population
growth. Gray curves indicate hole bands while black curves indicate
bulk particle bands.  Red and blue curves indicate edge states
propagating in opposite directions.  For clarity, edge states on only
one side of the system are plotted (on the opposite side, the roles of
the particle and hole edge bands are reversed).  Imaginary parts of
eigenvalues are given by dashed lines.
}
 \label{Fig:bands}
\end{figure}
We now arrive at the central result of this work. If we tune the quench parameter $\shift$ to satisfy
\begin{align}
\label{inst_cond}
\varepsilon_k^{(\nu)} -U_s \bar{n} < \shift <  \varepsilon_k^{(\nu)} +U_s \bar{n} \,,
\end{align}
the Bogoliubov energies $E_k^{(\nu)}$ becomes imaginary signifying a
dynamical instability for the $(k,\nu)$ mode. Physically, this
provides exponentially fast population of the unstable modes.
In other words, through (\ref{inst_cond}), the quenching
protocol gives a ``window'' (centered on $\Delta$ and of width
$2U_s \bar{n}$) of unstable modes in the spectrum $\varepsilon_k^{(\nu)}$.

To understand the instability criterion, we now 
discuss the single-particle eigenstates for the spin-$\pm1$ components.
%For sufficiently large systems in the strip geometry, the bulk energy modes are well 
%approximated by applying periodic boundary conditions along both directions.
As already mentioned, these are the eigenstates of the Haldane model
with opposite fluxes for the two spin components. The bulk states form
two bands of states separated by a band
gap of $6 \sqrt{3} \lambda$ centered about zero \cite{haldane88}.
In addition, the non-interacting spectrum exhibits topologically
protected edge modes which exist within the bulk gap.
In the Supplemental Material, we show that the dispersion of
the edge states is
\begin{align}
\varepsilon_{k}^{\rm edge} = \pm\frac{6w\lambda\sin(k)}{\sqrt{w^2+16\lambda^2 \sin^2(\frac{k}{2})}}
\end{align}
where spin-$\pm 1$ modes propagate in opposite directions.
If we tune $\Delta$ to sit inside of the band gap, then, for sufficiently small $U_s \bar{n}$, 
we can achieve the intriguing situation where the bulk states are stable while the edge states 
experience exponentially fast population growth.   A similar scheme for populating edge modes has been previously
reported in \cite{barnett13}.  Moreover, unlike \cite{barnett13}, due to the tuneability of $\Delta$,
the current scheme allows one to selectively populate states with particular momenta along the edge.
From (\ref{BogEn}), we see that the most unstable modes occur at momenta for which
$\varepsilon_{k}^{(\nu)} = \Delta$, so, for instance, when $\Delta = 0$, edge modes with momenta
$k=\pi$ will be populated most rapidly.
Bulk and edge bands are shown in Fig.\ \ref{Fig:bands} for two
particular quenches.
  
The symmetries (inversion, time reversal, and spin rotation) of the
Bogoliubov Hamiltonian greatly simplify the above analysis.  That is,
given the single particle energies of the Haldane model, the above
analysis essentially reduced to solving a two-mode problem.  For more
general couplings with less symmetry, it is easiest to proceed by
solving the Bogoliubov-de Gennes (BdG) equations.  
\begin{align}
\tau_3 H_k v_{k,\nu} = E_k^{(\nu)} v_{k,\nu}
\end{align}
where $H_k$ is the BdG Hamiltonian which, for our problem, can be
directly determined from (\ref{Eq:bog}).  For a system having length
$N |{\bf a}_1|$ along the direction with open boundary conditions,
$H_k$ is a ${2 \cal N} \times {2 \cal N}$ dimensional matrix where
${\cal N} = 6 N$ (the factor of 6 accounts for the spin and sublattice
degrees of freedom) while $\tau_3 = \sigma_3 \otimes
\mathds{1}_{\cal{N} \times \cal{N}}$.  The BdG Hamiltonian generically
possesses a `particle-hole' symmetry which requires the eigenvalues to
come in $\pm E_k^{(\nu)}$ pairs.  For each pair of stable (real)
eigenvalues, one member will have positive norm defined with the
$\tau_3$ metric, $v_{k,\nu^+}^\dagger \tau_3 v_{k,\nu^+} >0$, while
the other will have negative norm $v_{k,\nu^-}^\dagger \tau_3
v_{k,\nu^-} <0$.  In analogy with BCS superconductors, we refer to
bands composed of eigenstates with positive (negative) norms as
`particle' (`hole') bands, which are indicated in
Fig.~\ref{Fig:bands}.

For stable systems, positive (negative) norm states correspond to
positive (negative) eigenvalues $E_k^{(\nu)}$. This is not the case for
unstable systems.  In \cite{nakamura08}, the origin of a dynamical
instability was traced to positive and negative norm states that
become degenerate in the absence of pairing (non-particle number
conserving) terms.  Then, pairing terms generally lift such
degeneracies and lead to complex Bogoliubov energies.  This is
precisely the mechanism leading to the unstable edge modes in our
problem.  On the other hand, as is evident from
Fig.~\ref{Fig:bands}(b), one can have overlap between bulk particle
and bulk hole bands which do \emph{not} lead to dynamical
instabilities.

We interpret these bulk-band degeneracies as being protected by
symmetries of the problem.  Indeed, in the Supplemental Material we show that
small contributions to the Hamiltonian that break time-reversal,
inversion, or $S_z$ symmetry can hybridize the bulk particle and hole
bands leading to bulk dynamical instabilities which would obscure the population growth of the
edge modes.  Such symmetries can be used to construct other bosonic
models having unstable topological edge modes with stable bulk modes.
However, for definiteness and due to experimental relevance \cite{jotzu14}, we focus
on the interacting S1KM model in this work.

We now move on to discuss the physical consequences related to the quenching protocol.  
We first consider the number of particles excited into the spin-$\pm1$ modes as
a result of the quench.  Using (\ref{bilinear}), we find
\begin{align}
{\cal N}_{\pm 1}(t) = \!\!\!\!\!\!\sum_{i,m=\pm 1}\!\!\!\!
 \bra{\Psi (t)}\phih_{i,m}^\dagger \phih_{i,m} 
 \ket{\Psi (t)} %\notag \\
=2\sum_{k,\nu}\! |B_{k,\nu}|^2.
\end{align}
For quenches satisfying ($\ref{inst_cond}$) and chosen to select only edge
modes for instability, the population growth in ${\cal N}_{\pm 1}$ will be localized to the
edges of the system. Keeping only unstable modes and
linearizing the edge spectrum about $k=\pi$, for $|U_s| \bar{n}  t \gg 1$ one finds
\begin{equation}\label{popgrowth}
{\cal N}_{\pm 1}(t)  \approx \sqrt{1+\frac{16\lambda^2}{w^2}} \frac{N_2}{12 \lambda } \sqrt{\frac{ |U_s| \bar{n}}{\pi t}} e^{2 |U_s| \bar{n} t}
\end{equation}
where $N_2$ is the number of lattice sites along the ${\bf a}_2$ direction.
Note that this expression is also valid for negative $U_s$, 
\emph{e.g.}~for $^{87}$Rb atoms.

The quenching protocol is also expected to create a spin current along the edge.  
The continuity equation for the local spin moments $\Phihb_i^\dagger S_z  \Phihb_i$ gives an expression for
the spin current operator. At long wavelengths, the spin
current operator along the edge is found to be \cite{bernevig13}
\begin{equation}
\hat{{\cal J}}_k^{(s_z)} =\frac{1}{N_2} \sum_{k',m=\pm 1} 
m\,\Phihb_{k-\frac{k'}{2},m}^\dagger \partial_{k'} H_{k'}^{(m)} \Phihb_{k+\frac{k'}{2},m}
\end{equation}
where $\Phihb_{k,m}$ is a $2N_1$-dimensional vector composed
of annihilation operators for spin-$m$ bosons on sites in a unit cell
of the strip geometry. $H_{k}^{(m)}$
is the non-interacting matrix Bloch Hamiltonian for spin component $m$ which can be directly determined from (\ref{S1KM}).
We wish to evaluate the expectation value of this operator with the state $\ket{\Psi(t)}$.
Writing $\hat{{\cal J}}_k^{(s_z)}$ in an eigenbasis of the
non-interacting Hamiltonian, employing (\ref{bilinear}) and the
Feynman-Hellman relation, we find the intuitive relation
\begin{align}
J^{(s_z)}(t) \equiv \langle \hat{\cal J}^{(s_z)}_{k=0} \rangle = \frac{2}{N_2} \sum_{k',\nu}  \partial_{k'} \varepsilon_{k'}^{(\nu)} |B_{k',\nu}(t)|^2
\end{align}
where the two spin components have contributed an equal amount.  Additionally, the
 $k \ne 0$  components of 
$\langle \hat{{\cal J}}_k^{(s_z)} \rangle$ vanish.  Under the same
conditions as were used for the evaluation of ${\cal N}_{\pm 1}$, one finds
\begin{align}
\label{popgr}
J^{(s_z)}(t) \approx \frac{1}{4 } \sqrt{\frac{ |U_s| \bar{n}}{\pi t}} e^{2 |U_s| \bar{n} t}.
\end{align}

Before closing, we comment on the experimental feasibility of the
above protocol.  It is expected that quantum degeneracy in topological optical lattice
systems will be reached in the near future with advances in
experimental techniques.  This will open new doors
for exploring the non-equilibrium dynamics of topological band
systems.  In addition to the optical lattice, atoms are further
confined in experiments by an overall (typically harmonic) trapping
potential.  Harmonic traps can obscure the edge states in the
single-particle spectrum \cite{buchhold12} of such systems.  This
problem can be surmounted by using box-shaped traps 
which are now available \cite{gaunt13,chomaz15}.  On the other hand, 
%since
%the edge states we consider are in the excitation spectrum, 
we expect many aspects of our results with open boundary conditions to be
qualitatively correct for a harmonic trap provided the initial state
is in the Thomas-Fermi regime. The reason is that the condensate will screen
the trap potential, leading to an effective potential 
with sharp boundaries in the BdG equations for the edge
excitations of interest.  
Indeed, the dynamics of a spinor
condensate in a harmonic trap 
following a quench in the quadratic Zeeman field can be accurately
modeled by an effective spherical-box potential \cite{scherer10}. 
Due to the sharp boundaries of the effective potential, we expect that
well-defined edge states will continue to exist in this geometry.
Finally, we note
that our results will hold for either antiferromagnetic ($U_s>0$) or
ferromagnetic ($U_s<0$) interactions and so are relevant for both
$^{87}$Rb and $^{23}$Na condensates.

In summary, we have proposed a method whereby topological edge modes
are populated exponentially fast through a quantum quench.
Although edge states are typically unimportant for bosonic gases near
equilibrium, we have shown that the non-equilibrium 
dynamics after a quench can be dominated by
degrees of freedom localized on the boundary of the system. The growth
of the edge modes will be limited at longer times by interactions
not captured in the Bogoliubov theory. 
The long-time decay mechanism 
of the dynamically populated edge modes 
will also be due to these interaction terms.

This quenching protocol provides a means of collecting cold atoms
coherently in a quasi-one-dimensional structure without the need for
extra trapping lasers.  Though this work focused on ultracold atoms,
it will also be worthwhile to consider the parallels with photonic
lattices where pairing terms can be generated by nonlinear optical
methods.

We are grateful for support from the Schrodinger Scholarship Scheme at
Imperial College London (B.\ G.);  the European Union's
Seventh Framework Programme for research, technological
development, and demonstration under Grant No.~PCIG-GA-2013-631002  and the Aspen Center for Physics under
Grant No. PHYS-1066293 (R.\ B.).

\section{Supplemental Material}

\subsection{Derivation of the edge-state dispersion}

\begin{figure}
 \includegraphics[width=75mm,angle=0]{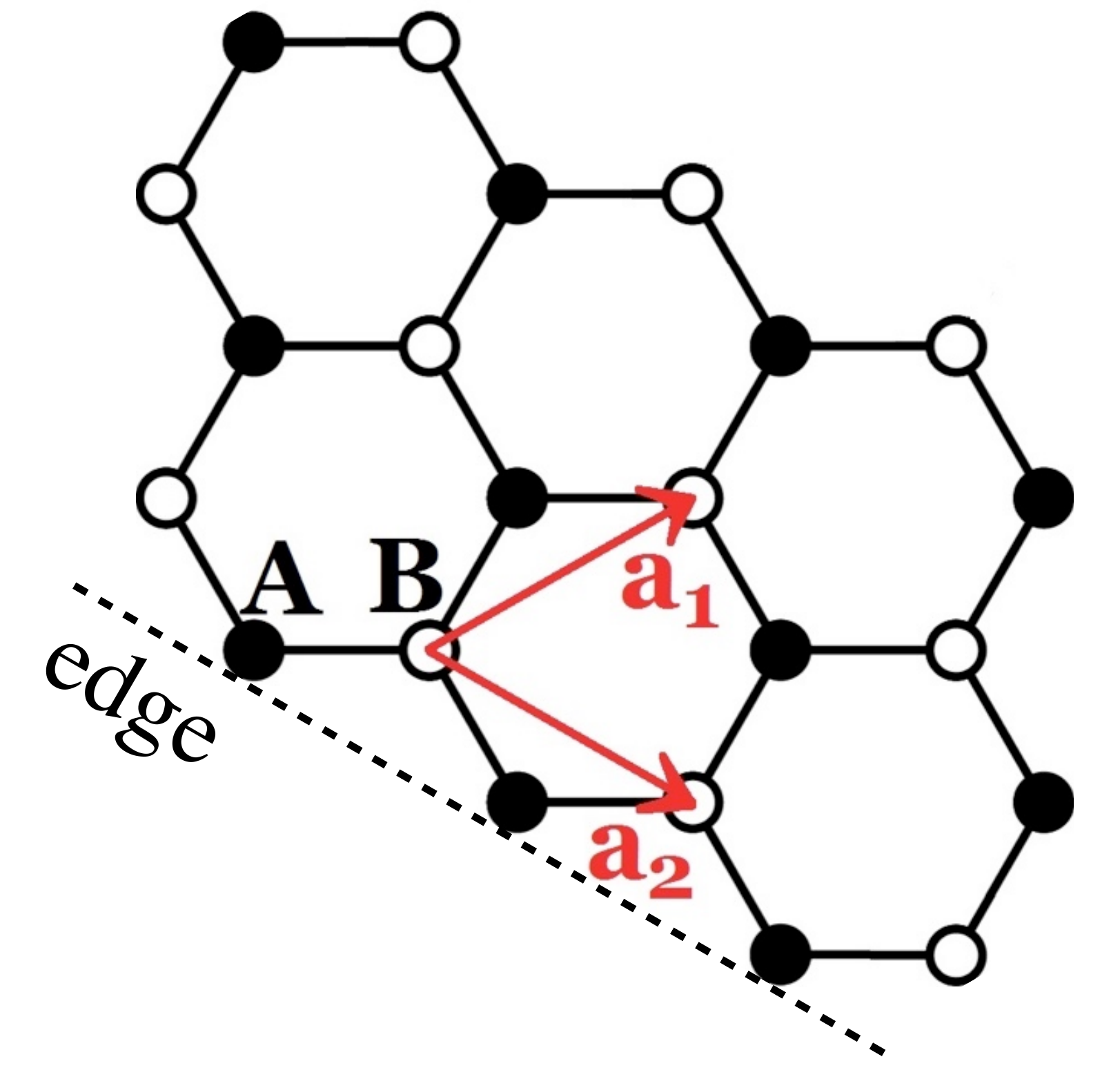}
 \caption{A portion of the hexagonal lattice used for the spin-one Kane-Mele Hamiltonian. Triangular sublattice sites
  $A$ and $B$ are labelled by closed and open circles respectively.  The lattice basis vectors are given by ${\bf a}_1$ and ${\bf a}_2$.}
 \label{Fig:lattice}
\end{figure}

Our aim is to analyze the single-particle edge states of $\Hh_{\rm S1KM}$ from the main text.  Note that since this
Hamiltonian is invariant under spin rotations about the $z$ axis, we can consider each spin component separately.
We first consider the Hamiltonian describing
$m=1$ spin component, which is the Haldane model.  
The primitive lattice vectors, as shown in Fig.\ \ref{Fig:lattice} are
$\latvec_1=\left(\frac{\sqrt{3}}{2},\frac{1}{2} \right)$ and
$\latvec_2=\left(\frac{\sqrt{3}}{2}, - \frac{1}{2} \right)$ 
where we have set the lattice constant to unity.
We impose periodic boundary conditions along the ${\bf a}_2$-direction and
take the system to be semi-infinite along the ${\bf a}_1$-direction. 
The single particle energies $\varepsilon_k$
are found by solving the following Schrodinger equation
\begin{align}
\label{SE}
R_k \psib_{n,k}+ V_k\psib_{n+1,k}+ V_k^\dagger \psib_{n-1,k}
=\varepsilon_k \psib_{n,k}
\end{align}
for $n > 1$ and
\begin{align}
R_k \psib_{n,k}+ V_k\psib_{n+1,k}
=\varepsilon_k \psib_{n,k}
\end{align}
for $n=1$.
Here $\psib_{n,k} = (\psi_{n,k}^{(A)}, \psi_{n,k}^{(B)})^T$ is a two-component vector where $n$
labels unit cell and $A, B$ label sublattice (see Fig.\ \ref{Fig:lattice}).    $R_k$ and $V_k$ are $2\times 2$
matrices given by
\begin{align}
R_k=
\left(
\begin{array}{cc}
-2\lambda \sin(k) & -w (1+e^{-ik}) \\ 
 -w (1+e^{ik})  & 2\lambda \sin(k) 
\end{array}
\right)
\end{align}
and
\begin{align}
V_k=
\left(
\begin{array}{cc}
-i\lambda (1-e^{-ik})& 0 \\ 
 -w & i\lambda (1-e^{- ik})
\end{array}
\right).
\end{align}
The Schrodinger equation written above and its constituent matrices directly follow from
the non-interacting Hamiltonian ${\Hh}_{\rm S1KM}$.

We search for a solution of the form 
\begin{align}
\psib_{n,k} = \frac{1}{\kappa_{1,k} - \kappa_{2,k}} \left[(\kappa_{1,k})^n - (\kappa_{2,k})^n) \right] \psib_{1,k}
\end{align}
where we require that $|\kappa_{1,k}| , |\kappa_{2,k}| <1$ so that the solution decays into the bulk.
Inserting this into the Schrodinger equation, we find
$
(R_k  + V_k \kappa_{1,k} +\frac{1}{\kappa_{1,k}} V_k^\dagger )\psib_{1,k} = \varepsilon_k \psib_{1,k}
$
and 
$
(R_k  + V_k \kappa_{2,k} +\frac{1}{\kappa_{2,k}} V_k^\dagger )\psib_{1,k} = \varepsilon_k \psib_{1,k}
$
which can be rearranged to give the following two equations:
\begin{align}
 V_k^{-1} V_k^\dagger \psib_{1,k} &= \kappa_{1,k} \kappa_{2,k} \psib_1, \\
 \left( R_k + (\kappa_{1,k}+\kappa_{2,k})V_k  \right) \psib_{1,k} &= \varepsilon_k \psib_{1,k}.
\end{align}
These equations can be directly solved to determine $\psib_1$ (up to a prefactor), $\kappa_1$, $\kappa_2$,
and $\varepsilon_k$.   In particular, for the edge dispersion we find
\begin{align}
\varepsilon_{k} = -\frac{6w\lambda\sin(k)}{\sqrt{w^2+16\lambda^2 \sin^2(\frac{k}{2})}}.
\end{align}
The edge mode energies for the $m=0$ and $m=-1$ spin components can be found with the
replacements $\lambda \rightarrow 0$ and $\lambda \rightarrow -\lambda$ respectively.  

\subsection{Symmetry protection of bulk instabilities}

\begin{figure*}
 \includegraphics[width=175mm,angle=0]{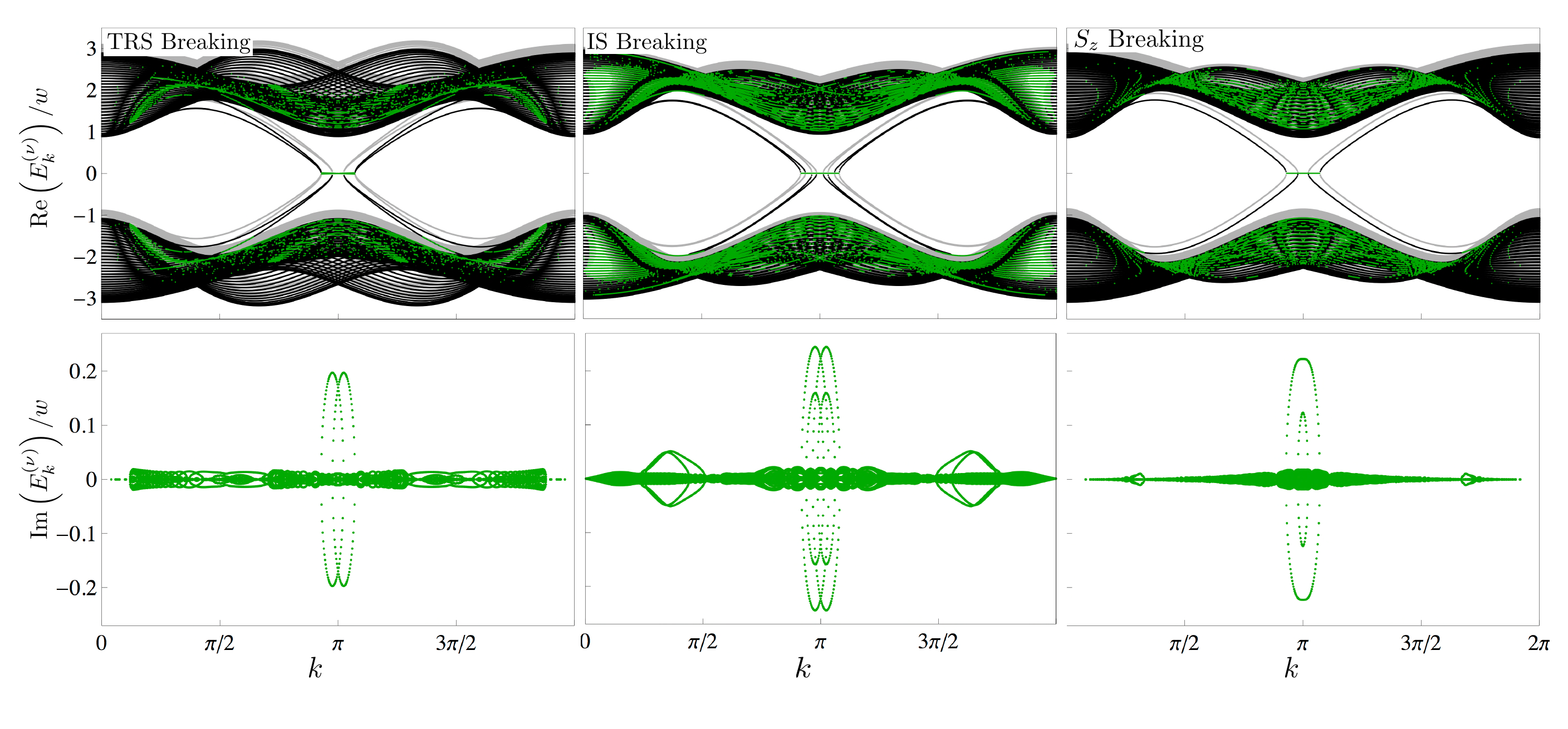}
\caption{
Bogoliubov energy spectrum $E_k^{(\nu)}$ for the interacting spin-one Kane-Mele model with  broken time-reversal, inversion, and $S_z$ symmetries. Parameters are chosen to be $U_s \bar n=0.2 w$,
$\lambda=w/2$,  $q_f=-3w-3U_s \bar{n}/2$, and $\delta =0.095w$. Gray curves indicate hole bands while black curves indicate bulk bands. 
The energy values corresponding to dynamically unstable modes (with finite imaginary component) are colored in green.  For clarity, modes from the zero spin component (which is always stable) are excluded from these plots.
}
\label{Fig:SymmBreaking}
\end{figure*}
We now show that small perturbations to the Hamiltonian $\Hh_B$  (Eq.~4 from the main text) that break time reversal symmetry (TRS), inversion symmetry (IS), or $S_z$ symmetry can lead to the emergence of dynamical instabilities coming from the hybridization of bulk modes in the particle-hole picture. 
Below we give three terms that we add to the Hamiltonian $\Hh_B$ which break either TRS, IS, or $S_z$ symmetry while preserving the other two.
To break $S_z$ symmetry we consider the contribution of same spin-component pairing terms
\begin{align}
\hat{V}_{S_z} = \delta\sum_i \left( \phih_{i,1}\phih_{i,1} +\phih_{i,-1}\phih_{i,-1} +{\rm H.c.} \right)
\end{align}
where $\delta$ is taken to be a real and small in comparison to other energy scales of the problem.
This term preserves TRS and IS.
Inversion symmetry can be broken, while keeping the TRS and $S_z$ symmetry intact, by adding a staggered pairing between the two triangular sublattices $A$ and $B$ (Fig.~\ref{Fig:lattice}):
\begin{align}
\hat{V}_I = \delta\sum_n \left( \phih_{n,1}^{(A)}\phih_{n,-1}^{(A)} -\phih_{n,1}^{(B)}\phih_{n,-1}^{(B)} + {\rm H.c.} \right)
\end{align}
where $n$ labels unit cells. % (unlike $j$ which ran over both sublattice sites). 
Finally, we consider a term that changes sign under time reversal, but preserves IS and  $S_z$ symmetry:
\begin{align}
\hat{V}_{T} = i\delta\sum_{\langle \langle ij \rangle \rangle} \nu_{ij} \phihb_i^\dagger S_z^2 \phihb_j
\end{align}
%We also consider a term that changes sign under time reversal, but remains invariant under sublattice interchange $(A\leftrightarrow B)$ and preserves $S_z$ symmetry:
%\begin{align}
%V_{T} = \delta\sum_{n} \left( \phih_{n,1}^{(A)\dagger}\phih_{n,1}^{(B)} -\phih_{n,-1}^{(A)\dagger}\phih_{n,-1}^{(B)} +{\rm h.c.} \right).
%\end{align}
In Fig.~\ref{Fig:SymmBreaking} we show the real and imaginary parts of Bogoliubov energy spectrum computed numerically after adding each of these terms to the Hamiltonian $\Hh_B$.  In addition to edge dynamical instabilities discussed in the main text,  some  bulk bands acquire a small imaginary part in the energy spectrum. This effect is amplified with the increase of $\delta$ and disappears in the absence of the symmetry breaking terms.

We now move on to discuss a slightly more stringent sufficient condition for the absence of bulk instabilities.  
We impose the condition of TRS and $S_z$ symmetry, and further require that the mean-field density 
(used to derive the Bogoliubov Hamiltonian)  is spatially uniform.  With these requirements, it can be shown that 
$(\tau_3 H_k)^2$ is a Hermitian matrix where $H_k$ is the BdG Hamiltonian introduced in Eq.~(11) from the main text.  This implies that the eigenvalues of 
$(\tau_3 H_k)^2$, namely
$(E_k^{(\nu)})^2$, will be real numbers and so $E_{k}^{(\nu)}$ will either be pure real or pure imaginary.
This means that an imaginary part in the energy spectrum will arise only in the region of overlap of particle and hole bands with vanishing real part of the energy. When the zero value of the real energy spectrum is situated in the gapped region, no dynamical instability is induced except in the crossing edge states. In this sense, 
these symmetries protect the bulk bands from becoming dynamically unstable.  While the model
we consider in the main text possesses these symmetries, it is possible that less stringent conditions exist.

\bibliographystyle{apsrev4-1} 
%\bibliography{S1KM}


\end{document}